\documentclass[prc,aps,a4paper,groupedaddress,superscriptaddress,nofootinbib,showpacs
,preprintnumbers,onecolumn]{revtex4}
\usepackage{graphicx}% Include figure files
\usepackage{amsfonts}
\usepackage{amssymb}
\usepackage{amsmath}
\usepackage{natbib}
\usepackage{dcolumn}% Align table columns on decimal point
\usepackage{bm} % bold math
\newcommand{\bwt}{\begin{widetext}}
\newcommand{\ewt}{\end{widetext}}
\newcommand{\beq}{\begin{equation}}
\newcommand{\eeq}{\end{equation}}
\newcommand{\bea}{\begin{eqnarray}}
\newcommand{\eea}{\end{eqnarray}}
\begin{document}
\title{Quark-hadron phase transition in a neutron star under strong magnetic fields}
\author{A.~Rabhi}
\email{rabhi@teor.fis.uc.pt}
\affiliation{Centro de F\' {\i}sica Computacional, Department of Physics, University of Coimbra, 3004-516 Coimbra, Portugal} 
\affiliation{Laboratoire de Physique de la Mati\`ere Condens\'ee,
Facult\'e des Sciences de Tunis, Campus Universitaire, Le Belv\'ed\`ere-1060, Tunisia}
\author{H.~Pais}
\email{pais.lena@gmail.com}
\affiliation{Centro de F\' {\i}sica Computacional, Department of Physics, University of Coimbra, 3004-516 Coimbra, Portugal}
\author{P.~K.~Panda}
\email{prafulla.k.panda@gmail.com}
\affiliation{Indian Association for the Cultivation of Sciences, Jadavpur, Kolkata-700 032, India}
\affiliation{Centro de F\' {\i}sica Computacional, Department of Physics, University of Coimbra, 3004-516 Coimbra, Portugal} 
\author{C.~Provid\^encia}
\email{cp@teor.fis.uc.pt}
\affiliation{Centro de F\' {\i}sica Computacional, Department of Physics, University of Coimbra, 3004-516 Coimbra, Portugal} 
\date{\today}    
\begin{abstract}
We study the effect of a strong magnetic field on the properties of neutron
stars with a quark-hadron phase transition. It is shown that the magnetic
field prevents the appearance of a quark phase, enhances the leptonic
fraction, decreases the baryonic density extension of the mixed phase and
stiffens the total equation of state, including both the stellar matter and the magnetic field
contributions. Two parametrisations of a density dependent static magnetic
field, increasing, respectively, fast and slowly with the density and reaching
$2-4\times 10^{18}$G in the center of the star, are considered. The compact
stars with strong magnetic fields have maximum mass configurations with larger masses and radius and smaller quark fractions. The parametrisation of the magnetic field with density has a strong influence on the star properties.
\end{abstract}
\pacs{26.60.+c, 12.39.Ba, 21.65.+f, 97.60.Jd} 
\maketitle

\section{Introduction}
Neutron stars with very strong magnetic fields of the order of
$10^{14}-10^{15}$ G are known as magnetars and they are believed to be the
sources of the intense gamma and X rays \cite{duncan,usov,pacz}.
If the magnetic field at the center of the star has the same magnitude as the one
measured at the  surface of a magnetar, then the influence of the magnetic
field on the star's mass and radius is negligible.
However, according to the scalar virial theorem  the
maximum interior field strength could be as large as  10$^{18}$ G for a
compact star \cite{shap83}. For a field of this magnitude or larger 
considerable effects will be induced on EOS.

It has been shown by several authors that the magnetic fields larger than
$B=5\times 10^{18}$ G  will affect the EOS of compact stars
\cite{chakrabarty96,broderick}. In particular field-theoretical descriptions
based on the non-linear Walecka model (NLWM) \cite{bb} were used. The
generalization of these works to density-dependent hadronic models \cite{aziz08}
or the inclusion of the scalar-isovector $\delta$ mesons \cite{wei}
have also been applied to the description of hadronic stars with
intense magnetic fields. Quark stars with strong magnetic fields have
been described within the MIT bag model \cite{chakrabarty96, hpais, aurora} 
and the SU(2) version of the Nambu--Jona--Lasinio model (NJL) \cite{njl}, an
effective  model which includes the  chiral symmetry, a symmetry of QCD.
 
Compact stars are complex systems which may contain in their core a pure
quark phase or  a non-homogenous mixed quark-hadron phase
\cite{glen00}. We would like to investigate whether the presence of a
magnetic field could affect the presence of a quark core. The influence
of strong magnetic fields on quark-hadron phase transition was first
discussed in Ref.~\cite{chakrabarty97}, using a Dirac-Hartree-Fock
approach within a mean-field approximation to describe both the hadronic
and quark phases. For the hadronic matter they took a system of protons,
neutrons and electrons and for the quark phase the MIT bag model with
one-gluon exchange. A very hard quark EOS was used  so that
the hybrid star did not have a quark core. They have concluded that compact
stars have a smaller maximum mass in the presence of strong magnetic
fields, a result that does not agree with other works where hadronic
stars \cite{prakash2} or quark stars \cite{njl} in the presence of
strong magnetic fields, have been studied.

In the present work we will study the effect of a strong magnetic field
on the deconfinement phase transition inside a compact star. We will
consider two EOS for the quark phase, a softer and a harder one, chosen
in such a way that the star will have at least a small quark core if no
magnetic field exists. For the hadronic phase we consider a standard
hadronic EOS, the GM1 parametrisation of the non-linear Walecka model
proposed in \cite{gm91}, and include the baryonic octet. The
largest magnetic field measured at the surface of a star is $\sim
10^{15}$ G but nothing is known about the magnitude of magnetic fields
in the interior of a star. We will consider a magnetic field whose
magnitude increases from a value of   $\sim 10^{15}$ at the surface to a
maximum value of  $\sim 4\times 10^{18}$ G at the center. In
particular, we want to analyse how the parametrisation of the field
affects the star properties, such as the radius and mass.

This work is organized as follows: we make a brief review of the
formalism used for the hadronic, mixed and quark phases. Next we discuss
the parametrisation used for the magnetic field and present and discuss
the results obtained for the equation of state of stellar matter and
the  mass/radius properties of the corresponding compact star families,
for several values of the maximum magnetic field. At the end we will
draw some conclusions.

\section{Hadron matter equations of state}
For the description of the EOS of neutron star matter, we employ a field-theoretical 
approach in which the baryons interact via the exchange of $\sigma-\omega-\rho$ 
mesons in the presence of a static magnetic field $B$ along the $z$-axis. 
The Lagrangian density of the NLW model (GM1) can be written as  
\beq
{\cal L}= \sum_{b}{\cal L}_{b} + {\cal L}_{m}+ \sum_{l}{\cal L}_{l}.
\label{lan}
\eeq
The baryons, leptons ($l$=$e$, $\mu$), and mesons ($\sigma$, $\omega$ and  $\rho$) 
Lagrangians are given by
\bwt
\bea
{\cal L}_{b}&=&\bar{\Psi}_{b}\left(i\gamma_{\mu}\partial^{\mu}-q_{b}\gamma_{\mu}A^{\mu}- 
m_{b}+g_{\sigma  b}\sigma
-g_{\omega  b}\gamma_{\mu}\omega^{\mu}-g_{\rho  b}\tau_{3_{b}}\gamma_{\mu}\rho^{\mu}
-\frac{1}{2}\mu_{N}\kappa_{b}\sigma_{\mu \nu} F^{\mu \nu}\right )\Psi_{b} \cr
{\cal L}_{l}&=& \bar{\psi}_{l}\left(i\gamma_{\mu}\partial^{\mu}-q_{l}\gamma_{\mu}A^{\mu}
-m_{l}\right )\psi_{l}  \cr
{\cal L}_{m}&=&\frac{1}{2}\partial_{\mu}\sigma \partial^{\mu}\sigma
-\frac{1}{2}m^{2}_{\sigma}\sigma^{2}-U\left(\sigma \right)
+\frac{1}{2}m^{2}_{\omega}\omega_{\mu}\omega^{\mu}
-\frac{1}{4}\Omega^{\mu \nu} \Omega_{\mu \nu}  \cr
&-&\frac{1}{4} F^{\mu \nu}F_{\mu \nu}
+\frac{1}{2}m^{2}_{\rho}\rho_{\mu}\rho^{\mu}-\frac{1}{4}  P^{\mu \nu}P_{\mu \nu}
\label{lagran}
\eea
\ewt
where $\Psi_{b}$ and $\psi_{l} $ are the baryon and lepton Dirac fields,
respectively. The index $b$ runs over the eight lightest baryons $n$, $p$,
$\Lambda$, $\Sigma^-$, $\Sigma^0$, $\Sigma^+$,  $\Xi^-$ and
$\Xi^0$. (Neglecting the baryonic decuplet including the  $\Omega^{-}$ and
the $\Delta$ quartet, which appear only at very high densities, does not affect our conclusions). 
The baryon mass and isospin projection are denoted by $m_{b}$ and $\tau_{3_b}$, respectively. 
The mesonic and electromagnetic field  tensors are given by their usual expressions: 
$\Omega_{\mu \nu}=\partial_{\mu}\omega_{\nu}-\partial_{\nu}\omega_{\mu}$, $P_{\mu \nu}=\partial_{\mu}\rho_{\nu}-\partial_{\nu}\rho_{\mu}$, and  
$F_{\mu \nu}=\partial_{\mu}A_{\nu}-\partial_{\nu}A_{\mu}$. The baryon anomalous magnetic moments (AMM) are introduced via the coupling of the baryons to the electromagnetic field tensor with $\sigma_{\mu \nu}=\frac{i}{2}\left[\gamma_{\mu}, \gamma_{\nu}\right] $ and strength $\kappa_{b}$. The electromagnetic field is assumed to be externally generated (and thus has no associated field equation), and only frozen-field configurations will be considered. The  interaction couplings are denoted by $g$, the electromagnetic couplings by $q$ and the baryons, mesons and leptons masses by $m$.  The scalar self-interaction is taken to be of the form
\beq
U\left(\sigma \right)=\frac{1}{3}bm_{n}\left(g_{\sigma N}\sigma \right)^3+ \frac{1}{4}c \left(g_{\sigma N}\sigma \right)^4.
\eeq
The parameters of the model are the nucleon mass $M=939$ MeV, the masses of mesons $m_\sigma$, $m_\omega$, $m_\rho$ and the coupling parameters.

The meson-hyperon couplings are assumed to be fixed fractions of the
meson-nucleon couplings, $g_{i H}=x_{i H} g_{i N}$, where for each meson $i$,
the values of $x_{i H}$ are assumed equal for all hyperons $H$. The values of
$x_{i H}$ are chosen to reproduce the binding energy of the $\Lambda$ at
nuclear saturation as suggested in \cite{gm91}, and given in Table \ref{table2}.
\begin{table}
\caption{Static properties of baryons considered in this study. The mass, charge of baryon and strange charge of species $b$ are denoted by $m_b$, $q_b$ and $q^b_s$, respectively. The baryonic magnetic moment is denoted by $\mu_b$. The baryonic AMM is given by $\kappa_b=(\mu_b/\mu_N-q_bm_p/m_b)$, where $\mu_N$ is  the nuclear magneton.}
\label{table1}
\begin{ruledtabular}
\begin{tabular}{ c c c c c c}
      b              &  $m_{b}$(MeV) &  $q_{b}(e)$ & $q^b_s$ & $\mu_{b}/\mu_{N}$ &  $\kappa_{b}$  \\
\hline
       p           & 938.27 & 1 & 0 &  2.97 & 1.79 \\
       n           & 939.56 & 0 & 0 & -1.91 &-1.91 \\
$\Lambda^0$   & 1115.7 & 0 & -1& -0.61 &-0.61 \\
$\Sigma^+$ & 1189.4 &  1 & -1 &  2.46 &  1.67 \\
$\Sigma^0$  & 1192.6 &  0 & -1 &  1.61 &  1.61 \\
$\Sigma^-$   & 1197.4 & -1 & -1 & -1.16 & -038 \\
$\Xi^0$        & 1314.8  &  0 & -2 & -1.25 & -1.25\\
$\Xi^-$        & 1321.3  &  -1 & -2 & -0.65 & 0.06\\
\end{tabular}
\end{ruledtabular}
\end{table}

From the Lagrangian density in Eq.~(\ref{lan}), we obtain the following meson field equations in the mean-field approximation 
\bea
m^{2}_{\sigma} \sigma  +\frac{\partial U\left(\sigma \right)}{\partial\sigma}&=&\sum_{b}g_{\sigma b}\rho^{s}_{b}=g_{\sigma N}\sum_{b}x_{\sigma b}\rho^{s}_{b} \label{mes1} \\
m^{2}_{\omega} \omega^{0} &=& \sum_{b}g_{\omega b}\rho^{v}_{b}=g_{\omega N}\sum_{b}x_{\omega b}\rho^{v}_{b} \label{mes2} \\
m^{2}_{\rho} \rho^{0} &=&\sum_{b}g_{\rho b}\tau_{3_{b}}\rho^{v}_{b}=g_{\rho N}\sum_{b}x_{\rho b}\tau_{3_{b}}\rho^{v}_{b} \label{mes3}
\eea
where $\sigma=\left\langle  \sigma \right\rangle,\; \omega^{0}=\left\langle \omega^{0} \right\rangle\;\hbox{and}\;\rho=  \left\langle\rho^{0} \right\rangle$ are the nonvanishing expectation values of the mesons fields in uniform matter.

The Dirac equations for baryons and leptons are, respectively, given by 
\bwt
\bea
\big[i\gamma_{\mu}\partial^{\mu}-q_{b}\gamma_{\mu}A^{\mu}-m^{*}_{b}
-\gamma_{0}\left(g_{\omega}\omega^{0}
+g_{\rho}\tau_{3_{b}}\rho^{0}\right) 
-\frac{1}{2}\mu_{N}\kappa_{b}\sigma_{\mu \nu} F^{\mu \nu}\big] \Psi_{b}&=&0 \label{MFbary}\\
\left(i\gamma_{\mu}\partial^{\mu}-q_{l}\gamma_{\mu}A^{\mu}-m_{l} \right) \psi_{l}&=&0 \label{MFlep}
\eea
\ewt
where the effective baryon masses are given by 
\beq
m^{*}_{b}=m_{b}-g_{\sigma}\sigma \label{effmass}
\eeq 
and $\rho^{s}_b$ and $\rho^{v}_b$ are the scalar number density and the vector
number density, respectively. For  charge-neutral, neutrino free,
$\beta$-equilibrated  matter, the following conditions are satisfied:
\bea
\mu_{p}&=&\mu_{\Sigma^+}=\mu_{n}-\mu_{e},\cr
\mu_{\Lambda} &=&\mu_{\Sigma^0}=\mu_{\Xi^0}=\mu_{n},\cr
\mu_{\Sigma^-} &=&\mu_{\Xi^-}=\mu_{n}+\mu_{e},\cr
\mu_{\mu}&=&\mu_{e}, \label{beta-e}
\eea
\beq
%\mu_{b}&=&\mu_{n}-q_{b}\mu_{e}\cr
\sum_{b}q_{b}\rho^{v}_{b}+\sum_{l}q_{l}\rho^{v}_{l}=0.
\label{zeroc}
\eeq

The energy spectra for charged baryons, neutral baryons  and leptons (electrons and muons) are, respectively, given by
\bwt
\bea
E^{b}_{\nu, s}&=& \sqrt{k^{2}_{z}+\left(\sqrt{m^{* 2}_{b}+2\nu |q_{b}|B}-s\mu_{N}\kappa_{b}B \right)
^{2}}+g_{\omega b} \omega^{0}+\tau_{3_{b}}g_{\rho b}\rho^{0} \label{enspc1}\\
E^{b}_{s}&=& \sqrt{k^{2}_{z}+\left(\sqrt{m^{* 2}_{b}+k^{2}_{x}+k^{2}_{y}}-s\mu_{N}\kappa_{b}B 
\right)^{2}}+g_{\omega b} \omega^{0}+\tau_{3_{b}}g_{\rho b}\rho^{0}\label{enspc2} \\
E^{l}_{\nu, s}&=& \sqrt{k^{2}_{z}+m_{l}^{2}+2\nu |q_{l}| B}\label{enspc3}
\eea
\ewt
where $\nu=n+\frac{1}{2}-sgn(q)\frac{s}{2}=0, 1, 2, \ldots$ enumerates the Landau levels (LL) of the fermions 
with electric charge $q$, the quantum number $s$ is $+1$ for spin up and $-1$ for spin down cases.

For the charged baryons, we introduce the effective mass under the effect of a
magnetic field 
\beq
\bar m^c_{b}=\sqrt{m^{* 2}_{b}+2\nu |q_{b}|B}-s\mu_{N}\kappa_{b}B
\label{mc}
\eeq
the expressions of the scalar and vector densities are, respectively, given by~\cite{broderick}
\bea
\rho^{s}_{b}&=&\frac{|q_{b}|Bm^{*}_{b}}{2\pi^{2}}\sum_{\nu=0}^{\nu_{\mbox{\small
      max}}}\sum_{s}
\frac{\bar m^c_{b}}
%\sqrt{m^{* 2}_{b}+2\nu |q_{b}|B}-s\mu_{N}\kappa_{b}B}
{\sqrt{m^{* 2}_{b}+2\nu |q_{b}|B}} \ln\left|\frac{k^{b}_{F,\nu,s}+E^{b}_{F}}
{\bar m^c_{b}
%\sqrt{m^{* 2}_{b}+2\nu |q_{b}|B}-s\mu_{N}\kappa_{b}B
} \right|, \cr
\rho^{v}_{b}&=&\frac{|q_{b}|B}{2\pi^{2}}\sum_{\nu=0}^{\nu_{\mbox{\small max}}}\sum_{s}k^{b}_{F,\nu,s}. 
\eea
where $k^{b}_{F, \nu, s}$ is the Fermi momenta of charged baryon $b$ with quantum numbers $\nu$ and $s$. The Fermi energy $E^{b}_{F}$ is related to the Fermi momenta $k^{b}_{F, \nu, s}$ by
\beq
\left(k^{b}_{F,\nu,s}\right)^{2}=\left(E^{b}_{F}\right)^{2}-
%\left[\sqrt{m^{* 2}_{b}+2\nu |q_{b}|B}-s\mu_{N}\kappa_{b}B\right] 
\left(\bar m^c_{b}\right)^{2},
\eeq
For the neutral baryons, there is no LL quantum number $\nu$, so the Fermi momenta is denoted by 
$k^{b}_{F, s}$, and  the Fermi energy $E^{b}_{F}$ is given by
\beq
\left(k^{b} _{F,s}\right)^{2}=\left(E^{b}_{F}\right)^{2}-\bar{m}^{2}_{b},
\eeq
with
\beq
\bar{m}_{b}=m^{*}_{b}-s\mu_{N}\kappa_{b}B.
\eeq
The scalar and vector densities of the neutral baryon $b$ are, respectively, given by
\bea
\rho^{s}_{b}&=&\frac{m^{*}_{b}}{4\pi^{2}}\sum_{s} \left[E^ {b}_{F}k^{b}_{F, s}-\bar{m}^{2}_{b}\ln\left|
\frac{k^{b}_{F,s}+E^{b}_{F}}{\bar{m}_{b}} \right|\right],  \cr
\rho^{v}_{b}&=&\frac{1}{2\pi^{2}}\sum_{s}\left[ \frac{1}{3}\left(k^{b}_{F, s}\right) ^{3}
-\frac{1}{2}s\mu_{N}\kappa_{b}B
\left(\bar{m}_{b}k^{b}_{F,s}+\left(E^{b}_{F}\right)^{2}\left(\arcsin\left( \frac{\bar{m}_{b}}
{E^{b}_{F}}\right) -\frac{\pi}{2} \right)  \right) \right]. 
\eea
The vector density for leptons is given by
\beq
\rho^{v}_{l}=\frac{|q_{l}|B}{2\pi^{2}}\sum_{\nu=0}^{\nu_{\hbox{\small max}}}\sum_{s}k^{l}_{F,\nu,s},
\eeq
where $k^{l}_{F, \nu, s}$ is the lepton Fermi momenta, which is related to the Fermi energy $E^{l}_{F}$ by
\beq
\left(k^{l}_{F,\nu,s}\right)^{2}=\left(E^{l}_{F}\right)^{2}-\bar{m}^{2}_{l}, \quad l=e, \mu,
\eeq
with $ \bar{m}_{l}^2=m^{2}_{l}+2\nu |q_{l}|B$. The summation in $\nu$ in the above expressions terminates at $\nu_{max}$, the largest value of $\nu$ for which the square of Fermi momenta of the particle is still positive and which corresponds to the closest integer from below defined by the ratio
$$\nu_{max}=\left[\frac{(E^l_F)^2-m_l^2}{2 |q_l|\, B}\right],\quad \mbox{leptons}$$
$$\nu_{max}=\left[\frac{(E^b_F+s\,\mu_N\ \kappa_b\, B)^2 - {m_b^*}^2}{2 |q_b|\, B}\right], \quad \mbox{charged baryons}.$$
The chemical potentials of baryons and leptons are defined as 
\bea
\mu_{b}&=& E^{b}_{F}+g_{\omega b}\omega^{0}+g_{\rho b}\tau_{3_{b}}\rho^{0} \\
\mu_{l} &=& E^{l}_{F}=\sqrt{\left(k^{l}_{F,\nu,s}\right)^{2}+m^{2}_{l}+2\nu |q_{l}| B}.
\eea

We solve the coupled Eqs.~(\ref{mes1})-(\ref{zeroc}) self-consistently at a given baryon density $\rho=\sum_{b}\rho^{v}_{b}$  in the presence of strong magnetic fields. The energy density of neutron star matter is given by 
\beq
\varepsilon_m=\sum_{b} \varepsilon_{b}+\sum_{l=e,\mu}\varepsilon_{l}+\frac{1}
{2}m^{2}_{\sigma}\sigma^{2}+U\left(\sigma \right) +\frac{1}{2}m^{2}_{\omega}\omega^{2}_{0}+\frac{1}{2}m^{2}_{\rho}\rho^{2}_{0},
\label{ener}
\eeq 
where the energy densities of charged baryons $\varepsilon_b^c$, neutral baryons
$\varepsilon_b^n$, and leptons  $\varepsilon_l$ have, respectively, the
following forms
\bea
\varepsilon^c_{b}&=&\frac{|q_{b}|B}{4\pi^ {2}}\sum_{\nu=0}^{\nu_{\mbox{\small max}}}\sum_{s}\left[k^{b}_{F,\nu,s}E^{b}_{F}+\left(\bar m^c_b \right) ^{2} 
\ln\left|\frac{k^{b}_{F,\nu,s}+E^{b}_{F}}{\bar m^c_b} \right|\right] ,\cr
\varepsilon^n_{b}&=&\frac{1}{4\pi^ {2}}\sum_{s}\bigg[\frac{1}{2}k^{b}_{F, s}\left(E^{b}_{F}\right)^{3}-\frac{2}
{3}s\mu_{N}\kappa_{b} B \left(E^{b}_{F}\right)^{3}\left(\arcsin\left(\frac{\bar{m}_{b}}{E^{b}_{F}}
  \right)-\frac{\pi}{2}\right)\cr
&-&\left(\frac{1}{3}s\mu_{N}\kappa_{b} B
+\frac{1}{4}\bar{m}_{b}\right) 
%\cr &&
\left(\bar{m}_{b}k^{b}_{F, s}E^{b}_{F}+\bar{m}^{3}_{b}\ln\left|\frac{k^{b}_{F,s}+E^{b}_{F}}{\bar{m}_{b}} 
\right|\right) \bigg]
% \quad \hbox{neutral baryons}
\cr
\varepsilon_{l}&=&\frac{|q_{l}|B}{4\pi^ {2}}\sum_{\nu=0}^{\nu_{\mbox{\small max}}}\sum_{s}\left[k^{l}_{F,\nu,s}E^{l}_{F}
+ \bar m_l^2
%\left(m^{2}_{l}+2\nu |q_{l}|B\right) 
\ln\left|\frac{k^{l}_{F,\nu,s}+E^{l}_{F}}{\bar m_l
%\sqrt{m^{2}_{l}+2\nu |q_{l}| B}
} \right|\right].
\eea
The pressure of neutron star matter is given by 
\beq
P_{m}=\sum_{i}\mu_{i}\rho^{v}_{i}-\varepsilon_{m}=\mu_{n}\sum_{b}\rho^{v}_{b}
-\varepsilon_{m},
\label{press}
\eeq
where the charge neutrality and $\beta$-equilibrium conditions are used to get the last equality. 
The total energy density and the total pressure of the system are given, by adding the corresponding contribution of the magnetic field, 
\bea
\varepsilon &=&\varepsilon_m+\frac{B^{2}}{2}, \label{ener1}\cr
P &=& P_m+\frac{B^{2}}{2}.
\label{press1}
\eea

\section{Quark matter equations of state}
The MIT Bag model has been extensively used to describe quark matter. 
In its simplest form, the quarks are considered to be free inside a Bag and 
the thermodynamic properties are derived from the Fermi gas model. 
In the presence of a strong magnetic field Eqs.~(\ref{press1}) are still valid with the
energy density and  pressure of quark stellar matter, and
quark density,  respectively, given by
\bwt
\bea
\varepsilon_m &=& \sum_{i=u,d,s,e}\left\lbrace \frac{|q_{i}|B}{4\pi^2}\sum^{\nu^i_{max}}_{\nu=0}g_{i}\left[\mu_i\sqrt{\mu^2_i-\left(M^{(i)}_{\nu}\right)^{2}}
+\left(M^{(i)}_{\nu}\right)^{2}\ln\left|\frac{\mu_i+\sqrt{\mu^2_i-\left(M^{(i)}_{\nu}\right)^{2}}}{\left(M^{(i)}_{\nu}\right)^{2}}\right|  \right] \right\rbrace 
+\hbox{Bag}, \cr
P_m&=& \sum_{i=u,d,s,e}\left\lbrace \frac{|q_{i}|B} {4\pi^2}\sum^{\nu^i_{max}}_{\nu=0}g_{i}\left[\mu_i\sqrt{\mu^2_i-\left(M^{(i)}_{\nu}\right)^{2}}-\left(M^{(i)}_{\nu}\right)^{2}\ln\left|\frac{\mu_i+\sqrt{\mu^2_i-\left(M^{(i)}_{\nu}\right)^{2}}}{\left(M^{(i)}_{\nu}\right)^{2}}\right|  \right] \right\rbrace -\hbox{Bag}, \cr
\rho_q &=& \sum_{i=u,d,s}\left\lbrace \frac{|q_{i}|B}{2\pi^2}\sum^{\nu^i_{max}}_{\nu=0}g_{i}\sqrt{\mu^2_i-\left(M^{(i)}_{\nu}\right)^{2}} \right\rbrace, 
\eea
\ewt
where $M^{(i)}_{\nu}=\sqrt{m^2_i+2\nu |q_i| B}$ and $\nu $ runs over  the allowed
LL, $g_i$ denotes the degeneracy of the i-th species, \textit{i.e.}, six for
the quarks and two for the leptons, $m_q$  the quarks masses and Bag represents the bag pressure.

We use $m_u = m_d = 5.5$MeV, $m_s = 150.0$MeV and two values for the Bag $(165\hbox{MeV})^4$ and $(180\hbox{MeV})^4$. 
In what follows, the baryon density $\rho$ is defined as $\rho = \sum_q \rho_q/3$.

In a star with quark matter we must impose both beta equilibrium and charge neutrality. 
For $\beta$-equilibrium matter we must add the contribution of the leptons as free Fermi 
gases (electrons and muons) to the energy density, pressure and entropy density. 
The relations between the chemical potentials of the different particles are given by
\beq
\mu_s=\mu_d=\mu_u+\mu_e, \quad \mu_e=\mu_\mu.
\eeq
In terms of neutron and electric charge chemical potentials $\mu_n$ and $\mu_e$, one has
\beq
\mu_u=\frac{1}{3}\mu_n-\frac{2}{3}\mu_e, \: \hbox{and} \: \mu_s=\mu_d=\frac{1}{3}\mu_n+\frac{1}{3}\mu_e.
\label{chem1}
\eeq
For the charge neutrality we impose
\beq
\rho_e+\rho_\mu=\frac{1}{3}\left( 2\rho_u-\rho_d-\rho_s\right). 
\label{chem2}
\eeq 

\section{Mixed phase and hybrid star properties}
We study the hadron-quark phase transition which may occur in the core of massive neutron
stars. We will use the Gibbs criteria and global charge conservation to describe the   mixed phase of hadronic and quark
matter  \cite{glen00},  which for
two conserved charges, the electric charge and the baryonic charge, are given by 
\bea
\mu_{\mbox{\footnotesize HP}, i}=\mu_{\mbox{\footnotesize QP}, i}&=&\mu_i, \; i=n, e, \cr
P_{\mbox{\footnotesize HP}}\left(\mu_{\mbox{\footnotesize HP}} \right) &=&P_{\mbox{\footnotesize QP}}\left(\mu_{\mbox{\footnotesize QP}} \right). 
\label{gibbs}
\eea
where the subscripts HP and QP denote, respectively, the hadronic and the
quark phase. Imposing the condition of  global conservation of the electric
and baryonic charges, \textit{i.e}, both hadronic and quark matter are allowed
to be separately charged and have different baryonic densities.  We impose  
\bea
\chi \rho^{\mbox{\footnotesize QP}}_c+\left(1-\chi \right) \rho^{\mbox{\footnotesize HP}}_c+\rho^{l}_c&=&0, \cr
\chi \rho^{\mbox{\footnotesize QP}}+\left(1-\chi \right) \rho^{\mbox{\footnotesize HP}}&=&\rho
\label{charge}
\eea
where $\chi $ is the volume fraction occupied by the quark phase and
$\left(1-\chi \right)$  the volume fraction occupied by the hadron phase,
$\rho^{i}_c$ ($\rho^{i}$) are the electric (baryonic) charge densities of
quark and hadron phases and $\rho^{l}_c$ is the lepton electric charge density. 
The total energy density in the mixed phase reads
\bea
\varepsilon &=& \chi \varepsilon^{\mbox{\footnotesize QP}}+\left(1-\chi \right) \varepsilon^{\mbox{\footnotesize HP}}+\varepsilon_{l}.
\eea
The mixed phase is charactrized by a value of $\chi$ which varies between 0, at the
onset of the mixed phase, and 1, at the onset of the quark matter. The
baryonic densities $\rho_1$ and $\rho_2$ given in Tables \ref{table3} and
\ref{table5}, correspond, respectively, to $\chi=0$ and $\chi=1$ and define the
onset of the mixed phase ($\rho_1$) and the onset of the quark phase ($\rho_2$). 
The mixed phase is determined by simultaneously imposing the Gibbs conditions (\ref{gibbs}),
charge conservation conditions (\ref{charge}), chemical equilibrium conditions 
(\ref{beta-e}), (\ref{zeroc}), (\ref{chem1}) and (\ref{chem2}), and the field
equations (\ref{mes1}-\ref{MFlep}).

However, we  should point out that the Gibbs construction is an approximation that although taking
correctly into account the existence of two charge conserving conditions does not take into
account surface effects and the Coulomb field \cite{vos03,maru07}.  A complete treatment of the mixed phase
requires the knowledge of the surface tension between the two phases which is not well
established and may have a value between 10-100 MeV/fm$^2$ \cite{maru07}. The Gibbs construction
gives results close to the ones obtained with the lower
value of the above  surface tension range, and is recovered for a zero surface tension. 
We could have also considered the Maxwell construction, for which only baryon number
conservation is considered, {\em i. e.} the first condition of Eq. (\ref{gibbs}) is only
satisfied for $i=n$. It has been shown in \cite{maru07} that the  Maxwell description of
the mixed phase gives a good description if the surface tension is very large.
Therefore,  for comparison we will also determine the mixed phase
using a Maxwell construction, see Tables \ref{maxwell180} and \ref{maxwell165}.

The EOS for the mixed phase are then constructed. Consequently, we can compute the properties
of the neutron star. In the present study we consider very strong magnetic fields which have a contribution to the total pressure at least as large as the contribution coming from matter. This
means that stars with non-spherical configurations should be considered~\cite{boc95,prakash2}.
We will, however, consider  spherical stars and use the Tolman-Oppenheimer-Volkoff (TOV)
equations for the structure of a static, spherically symmetric star to obtain the mass-radius
(M-R) relation of neutron star. With this approximation our results and, therefore, conclusions should
be taken with care.

\begin{table*}
\caption{The parameter set GM1 \cite{gm91}used in the calculation.}
\label{table2}
\begin{ruledtabular}
\begin{tabular}{ c c c c c c c c c c cc}
$\rho_{0}$& -B/A & &$g_{\sigma N}/m_{\sigma}$ &$g_{\omega N}/m_{\omega}$&$g_{\rho N}/m_{\rho}$&  &  &  & &\\
 ($fm^{-3}$) &(MeV)& $M^{*}/M$ & (fm) & (fm) & (fm) &$x_{\sigma H}$&$x_{\omega H}$ & $x_{\rho H}$ &b &c \\
\hline
0.153&16.30&0.70& 3.434 & 2.674 & 2.100 & 0.600 & 0.653 & 0.600 &0.002947 &-0.001070 \\
\end{tabular}
\end{ruledtabular}
\end{table*}

\section{Results and discussion}

\begin{figure}[htb]
\vspace{1.5cm}
\centering
\includegraphics[width=0.7\linewidth,angle=0]{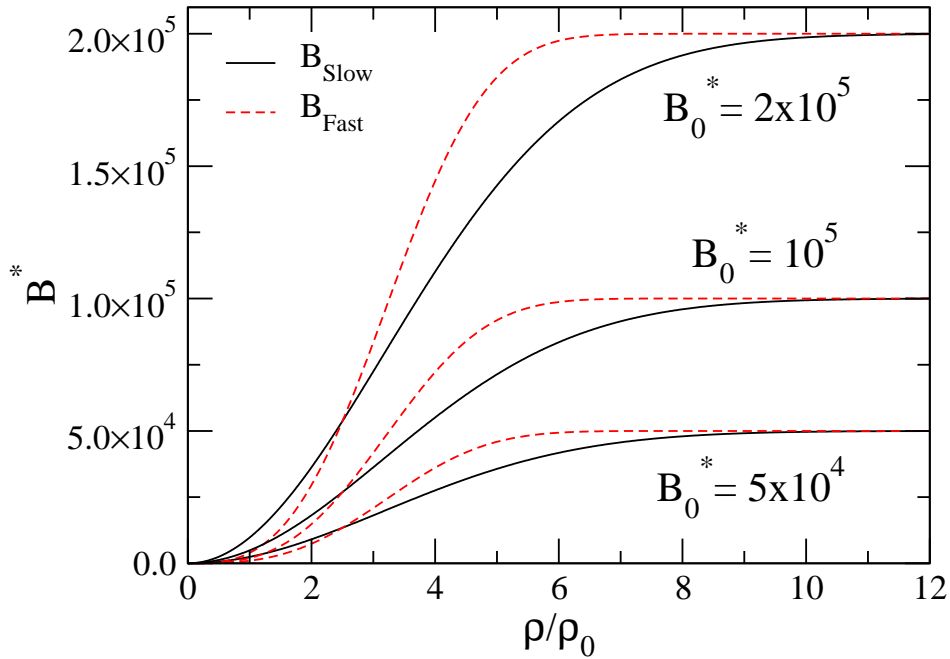}
\caption{(Color online). The magnetic field $B$ defined in (\ref{brho}) as
function of the baryonic density for several values of the parameter $B_{0}$
and for the slow ($\beta=0.05$ and $\gamma=2$), and fast parametrisations
($\beta=0.02$ and $\gamma=3$).}
\label{Bvar}
\end{figure}
 
In order to study the effect of strong magnetic fields on the structure of
neutron star  we use the GM1 parametrisation of the NLW model given in
Table.~\ref{table2}, \cite{gm91}. %It was shown in Ref.~\cite{Mao03} that 
 Since, to date, there is no
information available on the interior magnetic field of the star, we will
assume that the magnetic field is baryon density-dependent as suggested by
Ref.~\cite{chakrabarty97}. The variation of the magnetic field $B$ with the
baryons density $\rho$ from the center to the surface of a star is
parametrized~\cite{chakrabarty97, Mao03} by the following form
\beq
B\left(\frac{\rho}{\rho_0}\right) =B^{\hbox{surf}} + B_0\left[1-\exp\left\lbrace-\beta\left( \frac{\rho}{\rho_0}\right)^\gamma  \right\rbrace  \right],
\label{brho}
\eeq 
where $\rho_0$ is the saturation density, $B^{\hbox{surf}}$ is the magnetic
field at the surface taken equal to $10^{15}$G in accordance with the
values inferred from observations and $B_0$ represents the magnetic field
 for large densities. The parameters $\beta $ and $\gamma$ are
chosen in such way that the field decreases fast or slow with the density from
the centre to the surface.  In this work, we will use two sets of values: a
slowly  varying field with 
 $\beta=0.05$ and $\gamma=2$, and a fast varying one defined by  $\beta=0.02$
 and $\gamma=3$. We give the
magnetic field in units of the critical  field $B^c_e=4.414 \times 10^{13}$~G,
so that $B=B^* \, B^c_e$. 
We further take $B^*_0$ as a free parameter to check the effect of different
fields. 

In Fig.\ref{Bvar} we plot the variation of the magnetic field, for the
two parametrisations of the magnetic field as function of the baryonic density for several
values of the parameter $B_0^*$ in Eq.~(\ref{brho}). For the slowly varying field the
magnitude $B_0^*$ is reached for densities $\rho> 10 \rho_0$, where $\rho_0$ is
the saturation density, while for the fast changing field this value is
reached for $\rho> 5 \rho_0$. On the other hand, for densities $\rho< 3
\rho_0$ the slowly varying magnetic field is stronger. This fact will
influence the stiffness of the equation of state (EOS): the EOS is stiffer a)  at low densities if a slowly
varying field is considered; b) at high densities if a fast varying field is chosen.
\begin{figure}[tb]
\vspace{1.5cm}
\centering
\includegraphics[width=0.75\linewidth,angle=0]{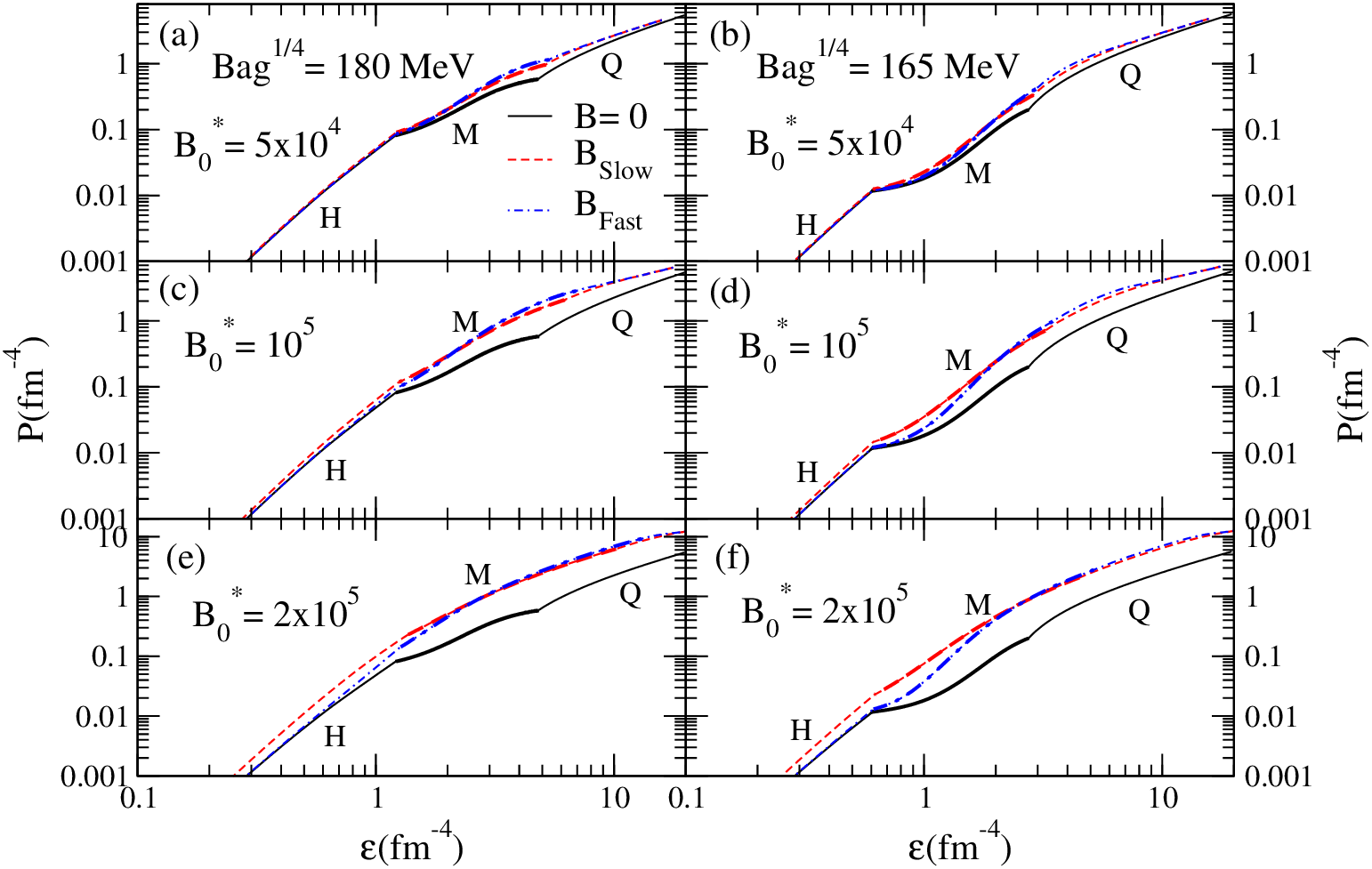}
\caption{(Color online). EOS for stellar matter for GM1  and for several
values of the magnetic field identified by the $B_0$ parameter of
Eq. (\ref{brho}). In the left panel (a), (c), and (e) for Bag$^{1/4}$=180
MeV and in the right panel (b), (d), and (f) for Bag$^{1/4}$=165 MeV. In
each figure we include the $B=0$ EOS, for reference, and the EOS obtained
with the slow and fast parametrisation of Eq. (\ref{brho}). The hadron,
mixed and quark phases are identified, respectevely, by a H, a M and thick
lines, and a  Q.}
\label{eos180}
\end{figure}
We present the numerical results for the  slowly and fast parametrisations of
the density-dependent magnetic field and for two Bag constants $(180\hbox{MeV})^{4}$ and
$(165\hbox{MeV})^{4}$.  In all figures where both cases are 
considered we show on the left panel the results for Bag$^{1/4}=$180 MeV 
and on the right panel the results for and Bag$^{1/4}$=165 MeV. 
Furthermore, in all the figures, we will only show  results obtained without
the inclusion of the baryonic AMM's, because for the intensity of the magnetic fields considered
the contribution to the EOS is negligible. The effect of the AMM is non
negligible for $B^*> 10^5$~\cite{broderick} but, as it will be shown later in this section
at the centre of the stars described in the present work these values will
never be reached.
\begin{figure}[tb]
\vspace{1.5cm}
\centering
\includegraphics[width=0.85\linewidth,angle=0]{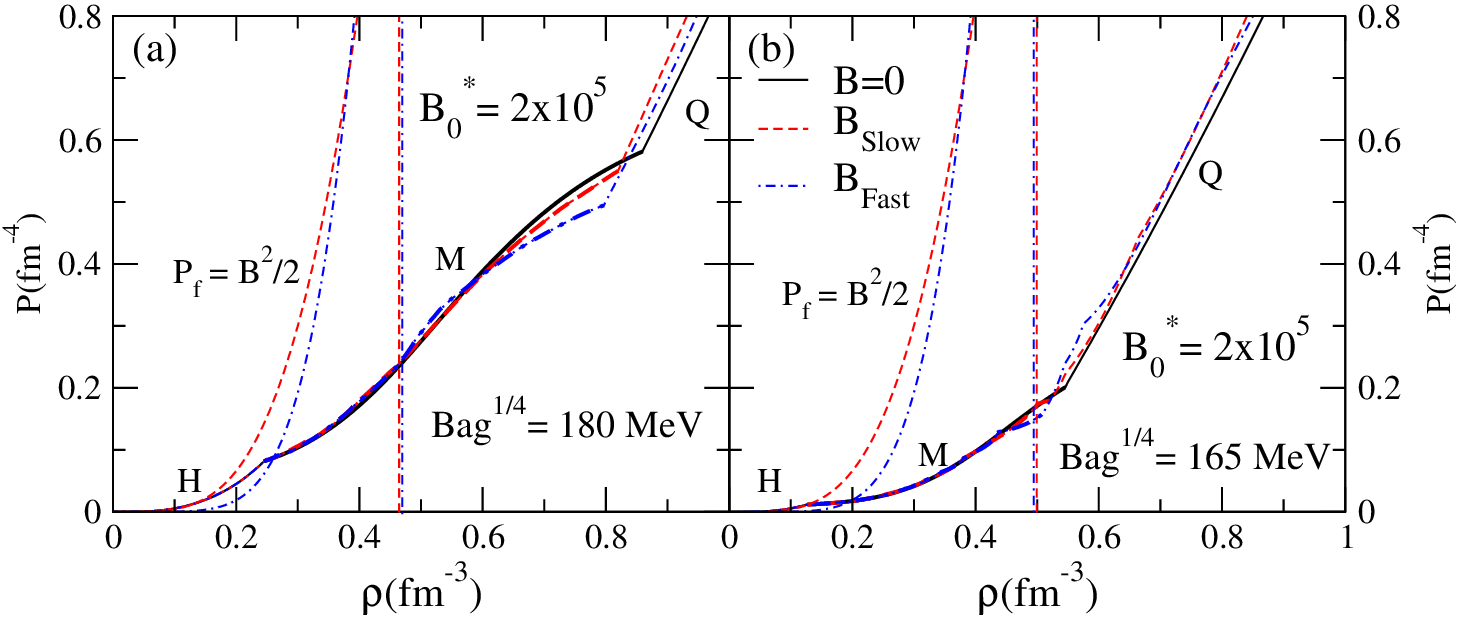}
\caption{(Color online). EOS for stellar matter for GM1 and for $B_0^*=2\times 10^5$. 
In the left panel (a) for Bag$^{1/4}$=180 MeV and in the right panel (b) for Bag$^{1/4}$=165 MeV. 
In each figure we include the $B=0$ EOS, for reference, and the EOS obtained with the slow and fast parametrisation of Eq. (\ref{brho}). The vertical lines identify the central baryonic density of the stable star with maximum mass.}
\label{eos180a}
\end{figure}
In Fig.~\ref{eos180}, we plot the EOS of neutron stellar matter, for the two
parametrisations of the magnetic field, defined by Eqs.~(\ref{press1}). For comparison we include in all
figures the EOS for $B=0$. The hadron phase is identified with a H (left part of the curves), the mixed phase is plotted with thick lines and identified with a M and the quark phase is identified with a Q (right part of the curves). We point out that these curves include both the contribution of stellar matter
and the contribution of the magnetic field in the energy density and the pressure.
In Fig.~\ref{eos180a} we plot only the pressure term  coming from the
stellar matter contribution, Eq.~(\ref{press}), as a function of the baryonic density for the largest $B^*_0=2\times 10^5$ in order to clearly show the effect of the magnetic field on the stellar matter contribution to the total EOS. We also plot, seperatly, the contribution of the magnetic field (thin line), which allows us to identify at which density the magnetic field contribution becomes larger than the matter contribution. The vertical lines identify the central density of the maxinmum mass star for both parameters bag considered.  Figs.~\ref{eos180} and~\ref{eos180a} are completed with Tables~\ref{table3}
and~\ref{table5}, where for the two parametrisations of the magnetic fields
and, respectively, for Bag$^{1/4}$=180 MeV and Bag$^{1/4}$=165 MeV,
the mixed phase density boundaries and the corresponding total energy 
densities defined in Eq.~(\ref{ener1}) are given for several values of the magnetic field. The
onset of the mixed phase occurs at density $\rho_1$, and the pure quark phase begins at density
$\rho_2$. In Tables \ref{maxwell180} and \ref{maxwell165} we give, for comparison, the same quantities obtained using  the Maxwell construction. Independently of the magnetic field, 
the onset  of the mixed phase and quark pure phase occurs at lower densities 
for the smaller value of bag pressure~\cite{dp03}, 
because  the EOS is much stiffer for
 $\hbox{Bag}^{1/4}=180 \hbox{ MeV}$ than $\hbox{Bag}^{1/4}=165 \hbox{ MeV}$. 
\begin{table*}[ht]
\caption{The phase density  boundaries, the  onset of the mixed phase $u_1=\rho_1/\rho_0$
and the onset of the pure quark phase $u_2=\rho_2/\rho_0$, and the corresponding total
energy densities, for several values of magnetic field using the two
parametrisations defined in Eq. (\ref{brho}) (slow and fast).  The bag constant is Bag$^{1/4}$=180 MeV, and the nuclear matter saturation density $ \rho_0=0.153\; \hbox{fm}^{-3}$ for GM1 model.} 
 \label{table3}
\begin{ruledtabular}
\begin{tabular}{cccccc}
$B^{*}_{0}$  &  & $u_1=\rho_1/\rho_0$ & $\varepsilon_1(\hbox{fm}^{-4}) $ & $u_2=\rho_2/\rho_0$ &$\varepsilon_2(\hbox{fm}^{-4})$  \\
\hline
$B = 0$                                      &         & 1.605 & 1.212 & 5.618 &  4.794  \\
$B^{*}_{0}=5\times10^{4}$  & slow & 1.604 & 1.220 & 5.588 &  5.145  \\
                                                   & fast  & 1.606 & 1.218 & 5.583 &  5.333   \\
$B^{*}_{0}=10^{5}$              & slow & 1.609 & 1.252 & 5.510 &  6.175   \\
                                                   & fast  & 1.605 & 1.228 & 5.477 &  6.914   \\
$B^{*}_{0}=2\times10^{5}$  & slow & 1.614 & 1.366 & 5.357 & 10.198  \\
                                                   & fast  & 1.603 & 1.273 & 5.206 & 13.017  \\
\end{tabular}
\end{ruledtabular}
\end{table*}

We now discuss  the effect of the presence of a strong magnetic field with
magnitude $B^*_0=5\times 10^4, \, 10^5,$ and   $2\times 10^5$ ($\sim
10^{18}-10^{19}$ G) on the equation of state (EOS). The main conclusions 
are: a) the presence of the field does not affect the baryonic density at the onset
of the mixed phase because the intensity of the field is too low ($B^*< 2\times 10^{4}$);
b) the baryonic density at the onset of the pure quark matter decreases with
the increase of the magnetic field, as seen in Fig.~\ref{eos180a} and discussed in
Ref.~\cite{chakrabarty97}. This is due to the fact that the EOS of quark matter under a strong magnetic
field becomes softer. The softening of the quark EOS with the increase of the
magnetic field has been obtained with other quark
models like the Nambu-Jona Lasinio (NJL) model \cite{klimenko03,njl};
c) due to the magnetic field contribution to the pressure and energy density,
Eqs. (\ref{ener1}) , the extension of the mixed phase in
terms of the energy density increases, as can be seen from Tables~\ref{table3}
and \ref{table5} and from Fig.~\ref{eos180},  mainly, 
shifting the onset of the quark phase to higher energy densities;
d) the contribution of the magnetic field makes the total EOS harder. However,
the slow and fast parametrisations have different effects.   At low densities the soft parametrisation correponds
to a stronger magnetic field and therefore gives rise to a harder EOS in that
range of densities  and favours the onset of the mixed phase at larger energy
densities. In Tables~\ref{table3} and~\ref{table5}  $\varepsilon_1$ is larger for the slow paramerisation.
 On the other hand, for large densities the fast
parametrisation  corresponds to larger magnetic fields and a harder EOS. The
onset of the quark phase occurs at larger energy densities, \textit{i.e.}  for a given $B_0^*$ value, $\varepsilon_2$ is larger for the fast parametrisation; 
e) the transition to the mixed phase is less affected by the magnetic field
than the transtion to a pure quark matter. This is due to the magnetic field
parametrisation, (\ref{brho}), which gives a weaker magnetic field at the
baryonic densities of the mixed phase onset. 
On the other hand, the transition to a pure quark phase is  strongly
affected  because for the densities at which it occurs
the magnetic field is stronger, more than 50\% (75\%) 
its maximum value for the slow (fast) varying magnetic field;
f) for all values of the  magnetic field considered,  the extension of  the
mixed-phase is larger for the larger bag pressure due to the larger stiffness
of the quark phase EOS. From Tables~\ref{maxwell180}
  and~\ref{maxwell165}, for the Maxwell construction similar
  conclusions are drawn, however with a larger baryonic density at the
  onset of the mixed phase, $\sim \rho_0$ larger for Bag$^{1/4}$=180 MeV and
  $\sim 0.5\rho_0$ larger for Bag$^{1/4}$=165 MeV, and a smaller one
  at the on set of the pure quark phase,  $\sim \rho_0$ smaller for Bag$^{1/4}$=180 MeV and
  $\sim 0.5\rho_0$ smaller for Bag$^{1/4}$=165 MeV. 

In Ref.~\cite{chakrabarty97} the magnetic field contribution was also taken 
into account in the EOS used to integrate the TOV equations, however this 
term enters with the wrong sign in the total pressure, it gives a negative 
contribution, and therefore, the EOS becomes very soft for strong magnetic 
fields and the  formation of very massive stars is not allowed.

\begin{table}[htb]
\caption{Same as Table~\ref{table3}, but for Bag$^{1/4}$=165 MeV.} 
\label{table5}
\begin{ruledtabular}
\begin{tabular}{cccccc}
$B^{*}_{0}$  & & $u_{1}=\rho_1/\rho_0$ & $\varepsilon_1(\hbox{fm}^{-4})$ & $u_{2}=\rho_2/\rho_0$ & $ \varepsilon_2(\hbox{fm}^{-4})$  \\
\hline
$B = 0$                                       &         & 0.832 & 0.613 & 3.567  & 2.748   \\
$B^{*}_{0} = 5\times10^{4}$ & slow & 0.831 & 0.613 & 3.546 & 2.869   \\
                                                    & fast  & 0.831 & 0.612 & 3.538 & 2.940   \\
$B^{*}_{0} = 10^{5}$             & slow  & 0.830 & 0.614 & 3.519 & 3.232  \\
                                                     & fast  & 0.831 & 0.612 & 3.478 & 3.467  \\
$B^{*}_{0} = 2\times10^{5}$  & slow & 0.830 & 0.623 & 3.402 & 4.496  \\
                                                     & fast   & 0.832 & 0.614 & 3.317 & 5.162  \\
\end{tabular}
\end{ruledtabular}
\end{table}

\begin{table*}[ht]
\caption{The phase density boundaries obtained with Maxwell construction, the  onset of the mixed phase $u_1=\rho_1/\rho_0$
and the onset of the pure quark phase $u_2=\rho_2/\rho_0$, and the corresponding total
energy densities, for several values of magnetic field using the two
parametrisations defined in Eq. (\ref{brho}) (slow and fast).  The bag constant is Bag$^{1/4}$=180 MeV, and the nuclear matter saturation density $ \rho_0=0.153\; \hbox{fm}^{-3}$ for GM1 model.} 
\label{maxwell180}
\begin{ruledtabular}
\begin{tabular}{cccccc}
$B^{*}_{0}$  &  & $u_1=\rho_1/\rho_0$ & $\varepsilon_1(\hbox{fm}^{-4}) $ & $u_2=\rho_2/\rho_0$ &$\varepsilon_2(\hbox{fm}^{-4})$  \\
\hline
$B = 0$                                      &         & 2.819 & 2.250 & 4.606  & 3.855  \\
$B^{*}_{0}=5\times10^{4}$  & slow & 2.799 & 2.297 & 4.580  & 4.089 \\
                                                   & fast  & 2.787 & 2.297 & 4.580 & 4.275 \\
$B^{*}_{0}=10^{5}$              & slow & 2.706 &  2.379 & 4.510  & 4.760  \\
                                                   & fast  & 2.715  & 2.422  & 4.471  & 5.420 \\  
$B^{*}_{0}=2\times10^{5}$  & slow & 2.477 & 2.628 & 4.228 & 6.918 \\ 
                                                   & fast  & 2.475 & 2.612 & 4.314 & 9.813 \\
\end{tabular}
\end{ruledtabular}
\end{table*}

\begin{figure}[b]
\vspace{1.5cm}
\centering
\includegraphics[width=0.9\linewidth,angle=0]{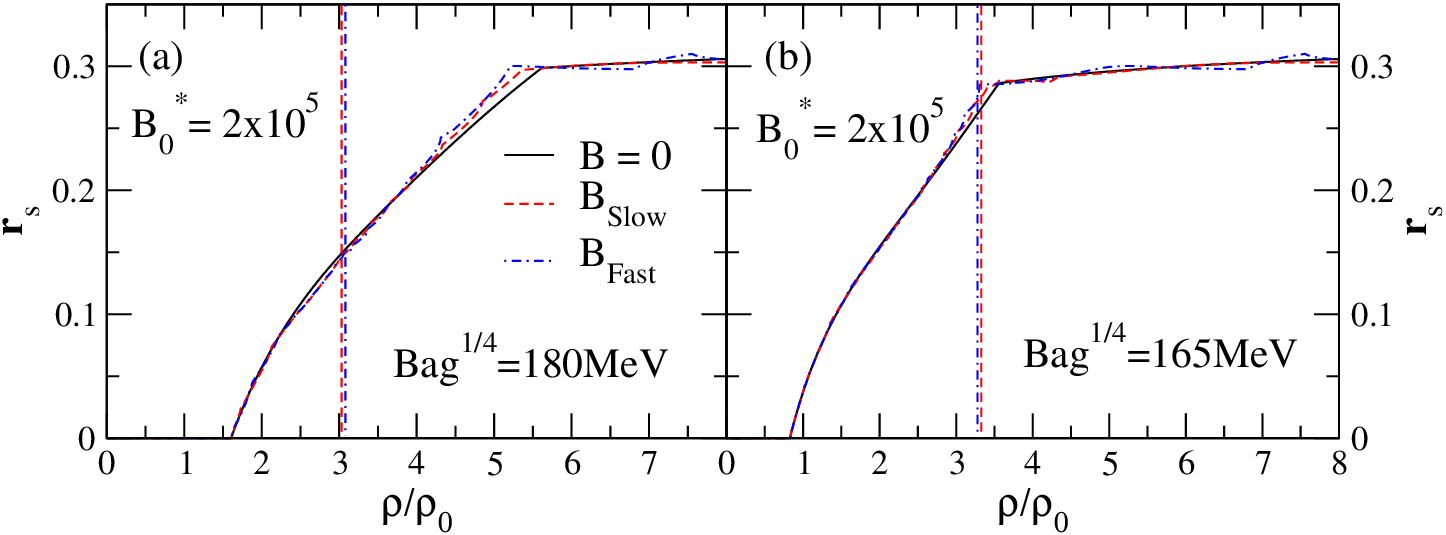}
\caption{(Color online). Strangeness fraction, defined in Eq. (\ref{strange}),
as function of the baryonic density, for $B^*_0=2\times 10^5$ in the
interior of the star, (a) for Bag$^{1/4}$=180 MeV and (b) for
Bag$^{1/4}$=165 MeV. The vertical lines identify the central baryonic
density of the stable star with maximum mass.}
\label{stra180}
\end{figure}

\begin{table}[htb]
\caption{Same as Table~\ref{maxwell180}, but for Bag$^{1/4}$=165 MeV.} 
\label{maxwell165}
\begin{ruledtabular}
\begin{tabular}{cccccc}
$B^{*}_{0}$  & & $u_{1}=\rho_1/\rho_0$ & $\varepsilon_1(\hbox{fm}^{-4})$ & $u_{2}=\rho_2/\rho_0$ & $ \varepsilon_2(\hbox{fm}^{-4})$  \\
\hline
$B = 0$                                       &         &  1.283 & 0.958 & 2.922 & 2.235 \\ 
$B^{*}_{0} = 5\times10^{4}$ & slow & 1.276 &  0.956 & 2.933 & 2.318 \\ 
                                                    & fast  & 1.276 &  0.953 & 2.933 & 2.339 \\ 
$B^{*}_{0} = 10^{5}$             & slow & 1.260 & 0.954 & 2.918 & 2.524 \\
                                                     & fast  &1.255  & 0.939 & 2.902 & 2.583 \\
$B^{*}_{0} = 2\times10^{5}$  & slow & 1.206 & 0.946 & 2.878 & 3.321 \\
                                                     & fast  & 1.196 & 0.901 & 2.847 & 3.507 \\
\end{tabular}
\end{ruledtabular}
\end{table}
As already discussed in Ref.~\cite{dp03}, the presence of strangeness in the
core of a neutron stars will affect some of their properties. In Fig.~\ref{stra180} we show the strangeness fraction defined by :
\beq
\mathfrak{r}_s=\chi \mathfrak{ r}^{QP}_s+(1-\chi) \mathfrak{r}^{HP}_s
\label{strange}
\eeq
with,
\beq
\mathfrak{r}^{QP}_s=\frac{\rho_s}{3\rho}, \qquad\mathfrak{r}^{HP}_s=\frac{\sum_{b}\left|q^{b}_ {s}\right|\rho_b}{3\rho}
\label{strange1}
\eeq
where $\rho_s$ is the strange quark density, and $q^b_s$ is the strange charge of baryon $b$, listed in Table~\ref{table1}.

Only the strongest field considered affects considerably the strangeness fraction.  At the onset of the pure quark phase, the strange quark fraction has reached $30\%$ 
of the baryonic matter. Although the fraction of strangeness varies in 
the mixed-phase, it has almost the same value in the quark phase. 
For $B_0^{*}= 2\times 10^5$, with both bag values and with the two
parametrisations of the magnetic field  used the effect of the Landau quantisation
appears, and the strangeness fraction obtained is slightly oscillating around
the $B=0$ result. The strangeness content increases slightly faster at larger densities with
the fast parametrisation. The presence of this strong magnetic field
shifts to lower densities the onset of the  $30\%$ fraction of strange matter,
corresponding to the onset of the quark phase. The thin vertical lines included in the Fig.~\ref{stra180}  identifies the central baryonic density of the maximum mass star. For Bag$^{1/4}$=180 MeV the strangeness fraction at given density is never larger than 15\%, while for Bag$^{1/4}$=165 MeV it
can be as large as 30\%.
     
\begin{table*}[htb]
\caption{Properties of the stable hybrid star with maximum mass, for Bag$^{1/4}$=180 MeV and for several
  values of magnetic field using [see  Eq. (\ref{brho})] with the  slow and fast
  parametrisations. $M_{max}$, $M^b_{max}$, R, $E_{0}$, $\rho^c$, and $B^*_c$
  are, respectively,  the gravitational and baryonic masses, the star radius,
  the central energy density, the central baryonic density, and the value of the magnetic field at the centre.} 
 \label{table4}
\begin{ruledtabular}
\begin{tabular}{cccccccc }
$B^{*}_{0}$ & & $M_{max} [M_\odot]$ & $M^b_{max} [M_\odot]$ & R [km] & $E_{0}[\hbox{fm}^{-4}]$ & $u^ c=\rho^c/\rho_0$ & $B^*_c $     \\
\hline
$B = 0$                                       &        & 1.42 & 1.57 & 10.22 & 8.05 & 8.680 &     \\
$B^{*}_{0} = 5\times 10^{4}$ & slow & 1.74 & 1.93 & 11.56 & 5.75 & 6.150 & 4.248 $\times10^{4}$ \\
                                                    & fast  & 1.85 & 2.06 & 11.49 & 5.85 & 6.092 & 4.948 $\times10^{4}$ \\
$B^{*}_{0} = 10^{5}$             & slow & 2.23 & 2.48 & 12.35 & 5.23 & 4.843 & 6.908 $\times10^{4}$ \\
                                                    & fast  & 2.33 & 2.55 & 11.88 & 5.62 & 4.595 &  8.565 $\times10^{4}$ \\
$B^{*}_{0} = 2\times 10^{5}$ & slow & 2.86 & 3.12 & 14.39 & 3.73 & 3.033 & 7.375 $\times10^{4}$ \\
                                                    & fast  & 2.76 & 2.88 & 13.15 & 4.41 & 3.092 & 8.926 $\times10^{4}$ \\
\end{tabular}
\end{ruledtabular}
\end{table*}

\begin{table*}[htb]
\caption{Same as Table~\ref{table4}, but for Bag$^{1/4}$=165 MeV.} 
\label{table6}
\begin{ruledtabular}
\begin{tabular}{cccccccc}
$B^{*}_{0}$ & & $M_{max} [M_\odot]$ & $M^{b}_{max} [M_\odot]$ & R [km] & $E_{0}[\hbox{fm}^{-4}]$ & $u^c=\rho^c/\rho_0 $ & $B^*_c $\\
\hline
$B = 0$                                        &        & 1.47 & 1.68 &  8.86 & 10.02 & 10.673 & \\
$B^{*}_{0} = 5\times10^{4}$  & slow & 1.72 & 1.95 &  9.80 & 8.24  & 8.588 & 4.877 $\times10^{4}$ \\
                                                     & fast  & 1.81 & 2.04 & 10.07 & 7.55 & 7.928 & 5.000 $\times10^{4}$ \\
$B^{*}_{0} = 10^{5}$              & slow & 2.14 & 2.36 & 11.07 & 6.52 & 5.902 & 8.251 $\times10^{4}$ \\
                                                     & fast  & 2.26 & 2.45 & 11.01 & 6.36 & 5.203 & 9.404 $\times10^{4}$ \\
$B^{*}_{0} = 2\times10^{5}$  & slow & 2.73 & 2.93 & 13.55 & 4.23 & 3.294 & 8.377 $\times10^{4}$ \\
                                                     & fast  & 2.68 & 2.78 & 12.50 & 4.80 & 3.222 & 9.759 $\times10^{4}$ \\
\end{tabular}
\end{ruledtabular}
\end{table*}

We next study the effect of the magnetic field on the properties of the stars
described by the EOS discussed above. We will consider that the stars have
spherical symmetry. 
The maximum gravitational and baryonic masses of the stable stars, radius, central
energy density, and central baryonic density, and the corresponding magnetic field at
the centre of the star,
 obtained by solving the
Tolman-Oppenheimer-Volkoff equation,  are given for Bag$^{1/4}$=180 MeV and
Bag$^{1/4}$=165 MeV in Tables~\ref{table4} and~\ref{table6},
respectively. We note that the central energy  and baryonic densities are
decreasing when the magnetic field increases, while the maximum masses and the corresponding
radius increase with the magnetic field, contrary to the results of
\cite{chakrabarty97} where the magnetic field contribution has entered with
the wrong sign in the total pressure expression, giving rise to very soft EOS
for large fields. 
This behavior is due to  the stiffening of
the EOS with  the magnetic field.
In the cases under study the  magnetic field at the
center of the maximum mass configuration is never larger than $4\times
10^{18}$ G a value close to the limit obtained from the virial theorem.  

\begin{figure}[htb]
\vspace{1.5cm}
\centering
\includegraphics[width=0.65\linewidth,angle=0]{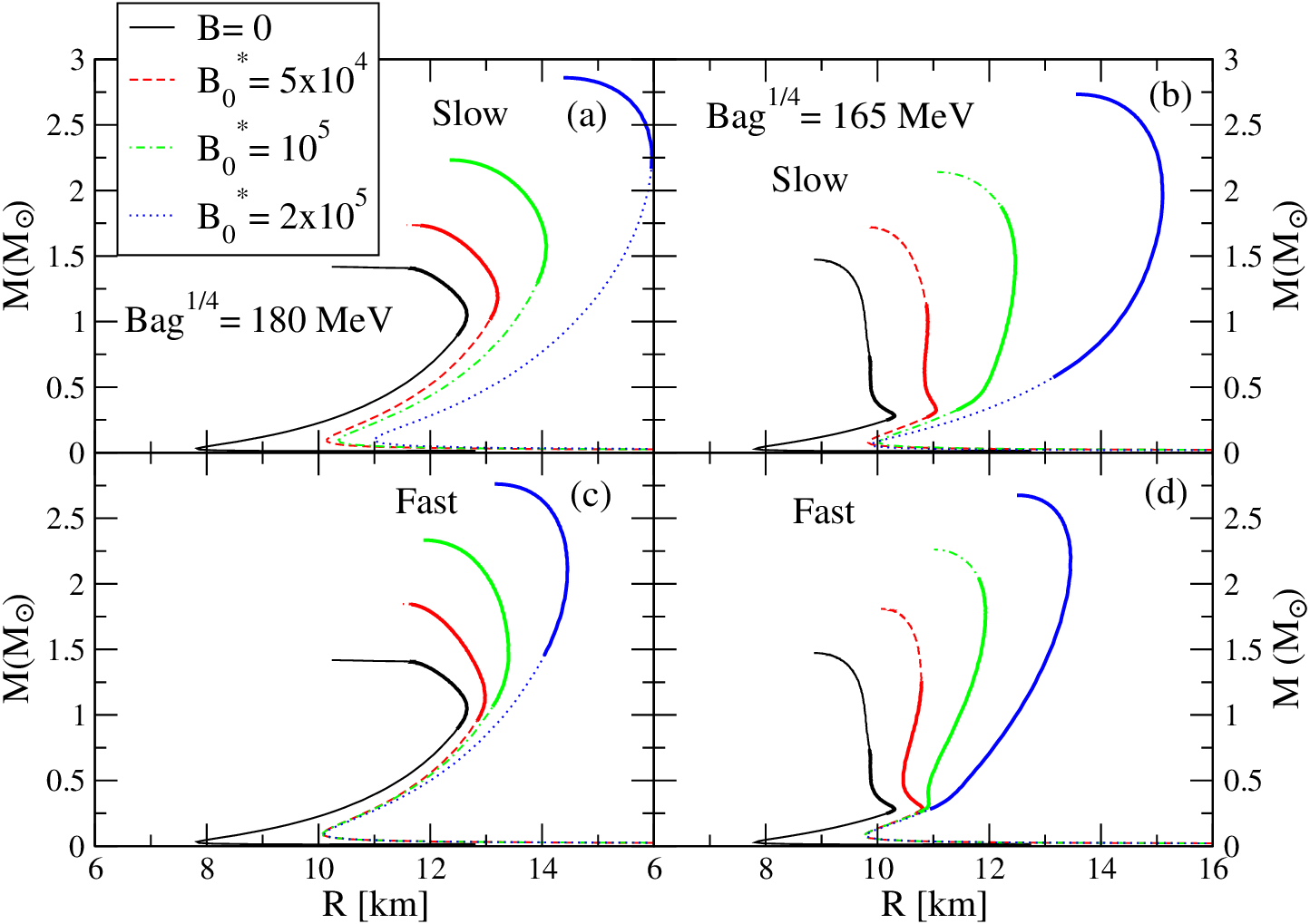}
\caption{(Color online). Mass-radius relation of the hybrid stars described in
  the present work for several
values of magnetic field using the slow and fast varying parametrisations
of $B$. (a) and (c) for Bag$^{1/4}$=180 MeV, and (b) and (d) for
Bag$^{1/4}$=165 MeV. The thick lines identify the stars with a mixed phase at
the center. Stars with smaller masses are hadronic stars with no quark matter,
and stars with larger masses have a pure quark core. For the $B_0^*=2\times
10^5$ there are no stars with a quark core.}
\label{massgrav}
\end{figure}

The mass/radius curves for the families of stars corresponding to the
maximum mass configurations given in Tables~\ref{table4} and~\ref{table6} are  plotted 
in Fig.~\ref{massgrav}. The thick lines correspond to stars for which the
central density lies within the  mixed phase. For the largest bag pressure
there is a quark core only for the no field configuration and the smallest
field considered. For the smallest bag pressure only the strongest field
considered gives rise to a no quark core configuration. It is interesting to notice the strong
effect of the magnetic field  parametrisation on the corresponding family star
properties. For  the stars that have no quark phase in their interior the central density is always smaller than $\rho\sim 2.5 \rho_0$ for which the slow and the fast varying parametrisations give the same magnetic field  magnitude, see Fig. \ref{Bvar}. Below this density the slowly varying parametrisation gives rise to larger magnetic fields and therefore to a harder EOS. The corresponding stars have larger radius and larger masses due to the larger incompressibility modulus. This explains why the maximum radius of stars with more than 0.5 $M_\odot$ is larger for the slowly varying field. This difference can be as large as 1 km for the largest field considered.  

However,  the largest mass configuration is obtained for central magnetic
fields larger for the fast varying field and therefore, the maximum mass is
larger in this case, except for the largest central value of the magnetic
field considered. For this value, the central density for the maximum mass
configuration is $\sim 3-3.3\,\rho_0$. From Fig. \ref{Bvar} it is clear that
these densities are just above the critical density values for which both
parametizations give the same magnetic field and, therefore, the star obtained
with the slowly varying parametrisation has, for most of the densities a
larger field. The size of the star is largely influenced by the lower density
layers and therefore most of the   maximum mass stars have a larger radius for
the slowly varying field, 
which give rise to larger magnetic fields in  the low density layers.

For a central magnetic field  $\sim 3\times\, 10^{18}$ G
we get maximum mass configuration with a mass 2.2-2.3 $M_\odot$. Slightly
larger central fields  predict even more massive stars with  $M>2.7\,
M_\odot$. These values would be able to describe highly massive compact stars,
such as the one associated to the millisecond pulsars PSR B1516 +
02B~\cite{freire1}, and the one in PSR J1748-2021B~\cite{freire2} in case they
are confirmed. Otherwise an  upper limit on the possible magnitude of the
magnetic field at the center of a compact star may be obtained.

\begin{figure}[htb]
\vspace{1.5cm}
\centering
\includegraphics[width=0.8\linewidth,angle=0]{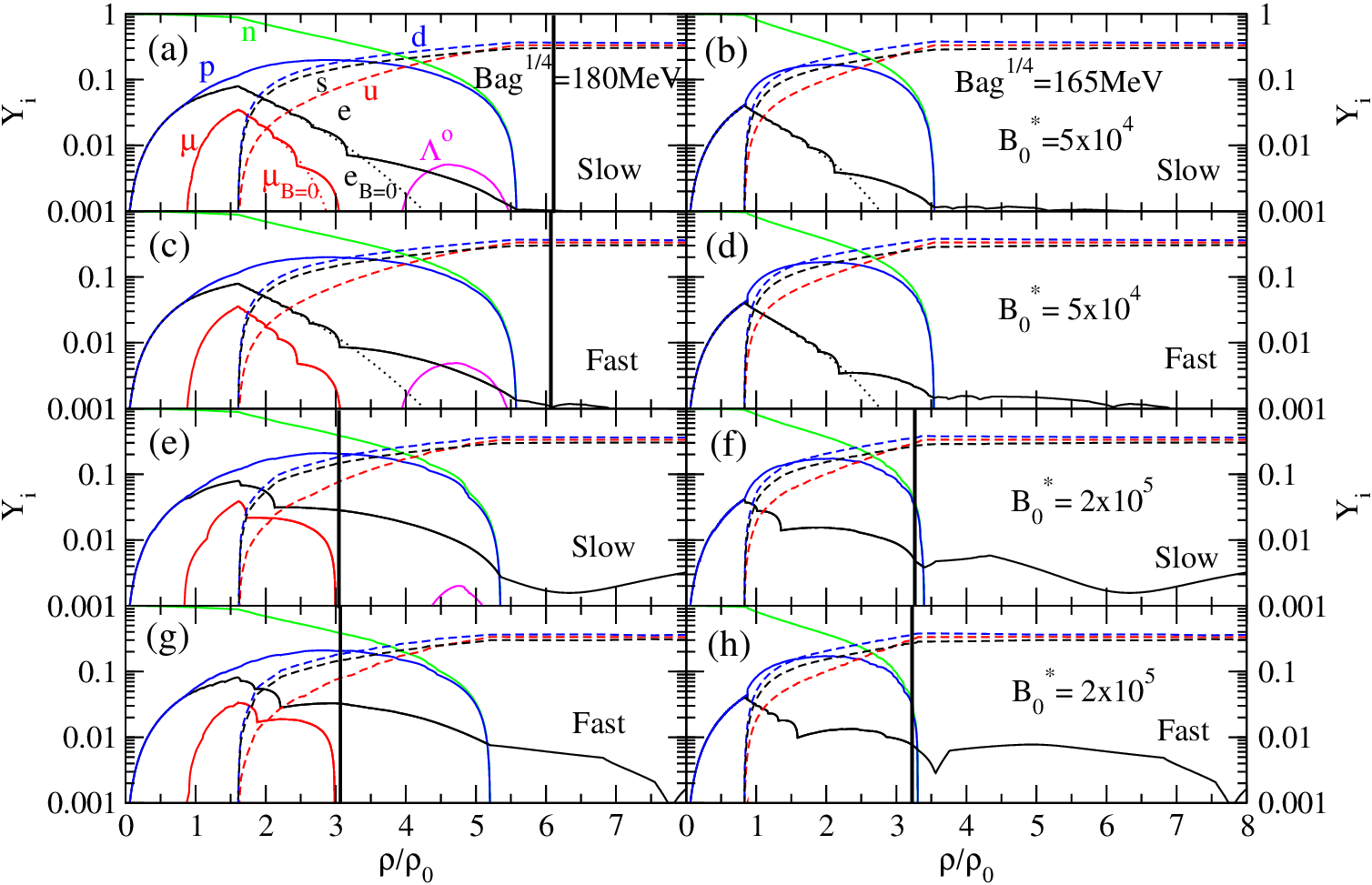}
\caption{(Color online). Particle fractions $\displaystyle Y_i=\rho_i/\rho$ as function of $\rho/\rho_0$, for i=baryons, leptons, and quarks, for two values of the magnetic field ($B^*_0=5\times 10^4$ (a)-(d), and $B^*_0=2\times 10^5$ (e)-(h)), with both parametrisations (slow and fast). We have considered two bag pressures: Bag$^{1/4}$=180 MeV for (a), (c), (e), and (g) figures, and Bag$^{1/4}$=165 MeV for (b), (d), (f), and (h) figures. The vertical lines represent the density $\rho^c$ at the center of the star with maximum mass.}
\label{frachq}
\end{figure}

The properties of the stars are strongly influenced by 
the magnitude of the magnetic field in their interior and the characteristics 
of the hadronic EOS become less important the larger the field is. It is also clear that a quark phase
is not favoured if a strong magnetic field exists in the interior of
the star. This conclusion is based on the Gibbs construction of the mixed phase. Within a Maxwell construction, the onset of  pure quark matter occurs at smaller densities and, therefore,  a small
quark core could still exist for the smaller Bag constant.

We have also looked at the baryonic and leptonic  composition of the stars.
In Fig.~\ref{frachq} we plot the lepton, baryon and the quark abundances.
We only show the particle fractions for the smallest and the largest magnetic 
fields considered in the present work. The vertical lines in the figures represent 
the central density of the configuration with maximum mass when it lies within 
the range of densities shown. Large magnetic fields prevent the appearance of 
a quark phase,  enhance the leptonic fraction and reduce slightly the mixed
phase baryonic density
extention. For strong magnetic fields, stars are mainly hadronic even when we 
consider a  soft quark EOS.  We also conclude that the only effect of the 
slow/fast varying magnetic field is the appearance at low/high densities of a larger 
leptonic fraction. The contribution of hyperons is 
minimal or even zero.

\section{Conclusions}

We have studied the influence of a static magnetic field on the deconfinement phase transition in the interior of a compact  star. For the hadronic phase we took the GM1 parametrisation of the NLWM \cite{gm91} and for the quark phase the MIT bag model with two different bag pressures, one corresponding to a soft and the other to a hard quark EOS. The mixed phase was built imposing the Gibbs conditions \cite{glen00}. 
It was considered that the magnitude of the magnetic field increases with density from a surface field of the order of $10^{15}$ G to a central field $\sim 5\times 10^{18}$G, and both a fast and a slow changing field were used. 

It was shown that due to the  weakness of the  magnetic field at low
densities, the onset of the mixed phase is not affected by the magnetic
field. The baryonic density at the onset of the pure quark phase is only
slightly reduced for the magnetic fields considered, but, due to the
contribution of the magnetic field to the total pressure and total energy density, the EOS of stellar matter in the presence of a strong magnetic field becomes much harder and the mixed phases extends to larger energy density ranges. As a consequence, the family of stars corresponding to these EOS have configurations with larger masses and radius. It was shown that the density dependence of the magnetic field will influence the star properties, with a slowly increasing field giving rise to larger radius.

Due to the magnetic field pressure,  the quark phase is not favoured, and
stars with strong magnetic field are mainly hadronic with very small hyperon
contributions. Stars with very high masses and radius are predicted and
maximum masses of observed compact stars may set an upper limit of the largest
possible magnetic field at the center of the star, $\sim 2\times 10^{18}$ G
for 2 $M_\odot$ stars. 

In the present work we have not taken into account the possibility of color
superconductivity (CS) which was shown to be important to describe the 
groundstate of quark matter at high chemical potential and  low temperature 
(for a review see~\cite{cs}). In particular, the effect of CS, in a color-flavor 
locked (CFL) phase, on the structure of hybrid stars was discussed in Ref.~\cite{cfl-ph}. 
Moreover, it has already been studied the effect of strong magnetic fields on the CS properties
which can be drastic, \cite{b-cs}. However, we point out that although a strong magnetic field may have important effects on the stellar quark matter EOS the contribution of the magnetic field energy to the total
energy of the star will impose severe restrictions on the maximum field a star may support and on the baryonic density reached at the centre of the star, \cite{prakash2}. In the present work, we have shown that: for very strong magnetic fields the central core of the star will probably never be in a quark phase,
but at most in a mixed phase.

Very strong magnetic fields can only occur in very young compact stars before
the magnetic field has decayed~\cite{magnetar}. We may therefore conclude that,
not only neutrino trapping in the protoneutron star phase~\cite{bombaci}, but  also a very strong magnetic field in the interior of a compact star strongly precludes the formation of a quark phase. Quark matter thermal nucleation in hot and dense hadronic matter has been proposed by several authors~\cite{nucleation}.
In these studies, it was found that the prompt formation of a critical size drop of quark matter via thermal activation  is possible above a temperature of about $2-3$ MeV, and as a consequence, it was inferred
that pure hadronic stars are converted to hybrid stars or quark stars within
the first seconds after their birth. The presence of a strong magnetic field would make this scenario less probable.

\begin{acknowledgments}
This work was partially supported by FEDER and Projects PTDC/FP/64707/2006 and CERN/FP/83505/2008, and  by COMPSTAR, an ESF Research Networking Programme.
\end{acknowledgments}


\begin{thebibliography}{34}

\bibitem{duncan} 
R. C. Duncan and C. Thompson, Astronphys. J. {\bf 392}, L9 (1992).

\bibitem{usov}
V. V. Usov, Nature 357, 472 (1992).

\bibitem{pacz}
B. Paczy\'nski, Acta Astron. 42, 145 (1992).

\bibitem{shap83} 
S. L. Shapiro and S. A. Teukolsky, {\em Black holes, white dwarfs and neutron stars}, Wiley-interscience New York, 1983.

\bibitem{chakrabarty96}
S. Chakrabarty, Phys. Rev. D 54, 1306 (1996).

\bibitem{broderick} 
A. Broderick, M. Prakash, and  J. M. Lattimer, Astrophys. J. 537, 351 (2000).

\bibitem{bb}
J. Boguta and A. R. Bodmer, Nucl. Phys. A292, 413 (1977);
B. D. Serot and J.D. Walecka, Adv. Nucl. Phys. 16, 1 (1986).

\bibitem{aziz08} 
A. Rabhi, C  Provid\^encia, and J. da  Provid\^encia, J. Phys. G: Nucl. Part. Phys. 35,125201 (2008).

\bibitem{wei} 
F.X. Wei, G.J. Mao, C.M. Ko, L.S. Kisslinger, H. Stocker and W. Greiner, J. Phys. G {\bf 32}, 47 (2006).

\bibitem{hpais}
H. Pais, Master thesis, University of Coimbra, 2008. 

\bibitem{aurora} 
R. Gonzalez Felipe, A. Perez Martinez, H. Perez Rojas and M. Orsaria, Phys. Rev. C {\bf  77}, 015807  (2008).

\bibitem{njl} 
D. P. Menezes, M. Benghi Pinto, S. S. Avancini, A. P\'erez
Mart\'{\i}nez, and C. Provid\^encia, Phys. Rev. C 79, 035807 (2009).

\bibitem{glen00} 
N. K. Glendenning,  {\em Compact stars}, Springer-Verlag New York , 2000.

\bibitem{chakrabarty97}
D. Bandyopadhyay, S. Chakrabarty, and S. Pal, Phys. Rev. Lett. 79, 2176 (1997).

\bibitem{prakash2}  
C. Y. Cardall, M. Prakash, and  J. M. Lattimer, Astrophys. J. 554, 322 (2001); 
A. Broderick, M. Prakash and J.M. Lattimer, Phys. Lett. B 531, 167 (2002).
\bibitem{gm91} 
N. K. Glendenning and S. A.  Moszkowski, Phys. Rev Lett. 67, 2414 (1991).

\bibitem{vos03} D. N. Voskresensky, M. Yasuhira, and T. Tatsumi, Phys. Lett. B 541, 93 (2002); D. N. Voskresensky, M. Yasuhira, and T. Tatsumi, Nucl. Phys. A723, 291 (2003).

\bibitem{maru07} 
T. Maruyama, S. Chiba, H-J Schulze, and T. Tatsumi, Phys. Rev. D 76, 123015 (2007).

\bibitem{boc95} 
M. Bocquet, S. Bonazzola, E. Gourgoulhon, and J. Novak, Astron. Astrophys. 301, 757 (1995).  

\bibitem{Mao03}
Guang-Jun Mao, Akira Iwamoto, and Zhu-Xia Li, Chin. J. Astron. Astrophys. Vol. 3, No. 4, 359-374 (2003).

\bibitem{dp03}  
D. P. Menezes and C. Provid\^encia, Phys. Rev. C 68, 035804 (2003); I. Bombaci, I. Parenti, I. Vida\~na, Astrophys. Jour. 614, 314 (2004).

\bibitem{klimenko03} D. Ebert, K.G. Klimenko, Nucl. Phys. {\bf A 728}, 203
  (2003).

\bibitem{freire1}
Paulo C. C. Freire, Alex Wolszczan, Maureen van den Berg, and Jason W. T. Hessels, Astrophys. J. 679,  1433 (2008).

\bibitem{freire2}  Paulo C. C. Freire, Scott M. Ransom, Steve Bégin, Ingrid H. Stairs, Jason W. T. Hessels, Lucille  Frey, Fernando Camilo, Astrophys. J. 675,  670 (2008).
\bibitem{cs} 
D. Bailin and A. Love, Phys. Rep. 107, 325 (1984);
K. Rajagopal and F. Wilczek, arXiv:hep-ph/0011333v2; 
M. Alford, Ann. Rev . Nucl. Part. Sci. 51, 131 (2001); 
Riv . Nuovo Cim. 25N3, 1 (2002); 
T. Schaefer, arXiv:hep-ph/0304281v2; 
D.H. Rischke, Prog. Part. Nucl. Phys. 52, 197 (2004);
Hai-cang Ren, arXiv:hep-ph/0404074v3.

\bibitem{cfl-ph}
M. Alford and S. Reddy, Phys. Rev. D 67, 074024 (2003);
P.~K.~Panda, D.~P.~Menezes, and C.~Provid\^encia, Phys. Rev. C 69, 025207 (2004);
P.~K.~Panda, D.~P.~Menezes, and C.~Provid\^encia, Phys. Rev. C 69, 058801 (2004).
 
\bibitem{b-cs} 
K. Iida and G. Baym, Phys. Rev. D 66, 014015 (2002); 
E. J. Ferrer, V. de la Incera and C. Manuel, Phys. Rev. Lett. 95, 152002 (2005); 
E. J. Ferrer, V. de la Incera and C. Manuel, Nucl.Phys. B 747, 88 (2006); 
E. J. Ferrer and V. de la Incera, Phys. Rev. Lett. 97, 122301 (2006); 
E. J. Ferrer and V. de la Incera, Phys. Rev. D 76, 045011 (2007); 
E. J. Ferrer and V. de la Incera, Phys. Rev. D 76, 114012 (2007) ; 
J. L. Noronha and I. A. Shovkovy, Phys. Rev. D 76, 105030 (2007); 
K. Fukushima and H. J. Warringa, Phys. Rev .Lett. 100, 032007 (2008);
D. M. Sedrakian, K. M. Shahabasyan, D. Blaschke and M. K. Shahabasyan, Astrophysics, Vol. 51, No. 4, (2008).

\bibitem{magnetar} http://www.physics.mcgill.ca/\verb ~ pulsar/magnetar/main.html

\bibitem{bombaci} I. Bombaci, I. Parenti and I. Vida\~na, Astrophys. J. {\bf 614}, 314 (2004); 
Ignazio Bombaci, Prafulla K. Panda, Constan\c ca Provid\^encia, and Isaac Vida\~na,
Phys. Rev. D 77, 083002 (2008).
 
\bibitem{nucleation} J. E. Horvath, O. G. Benvenuto and H. Vucetich, Phys. Rev. D 
 {\bf 45}, 3865 (1992);  J. E. Horvath, Phys. Rev. D {\bf 49}, 5590  (1994); 
M. L. Olesen and J. Madsen, Phys. Rev. D {\bf 49}, 2698 (1994).


\end{thebibliography}
\end{document}